\documentclass[aps]{revtex4}
\usepackage{amsmath}

\def\ddn{detection }
\def\dd{detector }
\def\dds{detectors }
\def\ddsn{detectors}
\def\t{temperature }
\def\tn{temperature}
\def\p{particle }
\def\pn{particle}
\def\ps{particles }
\def\psn{particles}
\def\zz{neutrino }

\def\zzs{neutrinos }

\def\be{\begin{equation}}
\def\ee{\end{equation}}
\def\bea{\begin{eqnarray}}
\def\eea{\end{eqnarray}}

\def\simlt{\lower.5ex\hbox{$\; \buildrel < \over \sim \;$}}
\def\simgt{\lower.5ex\hbox{$\; \buildrel > \over \sim \;$}}
\def\simpropto{\lower.2ex\hbox{$\; \buildrel \propto \over \sim
\;$}}

\begin{document}

\title{Bursts from the very early  universe }

\author{J. Silk}

\affiliation{Department of Physics, University of Oxford, Oxford
OX1
3RH}

\author{L. Stodolsky}

\affiliation{Max-Planck-Institut f\"ur Physik, F\"ohringer Ring 6,
80805 Munich}

\begin{abstract}
{ Bursts  of weakly interacting
particles  such as \zzs or
even more weakly interacting particles such as wimps and gravitons
  from the very early universe  would
offer  a much
deeper  ``look
back time'' to early epochs  than is possible with photons.
 We consider some of the
issues related to  the existence of such bursts  and their
detectability. Characterizing the burst rate by a probability $\cal
P$ per Hubble four-volume we find, for events in the
radiation-dominated era, that the natural unit of description is
the present intensity of the CMB times $\cal P$.  The
existence of such
bursts would make the
observation of  phenomena associated with very early times in
cosmology
   at least conceptually
possible.   One might even hope to probe   the transplanckian epoch
if
complexes  more weakly interacting than the graviton can exist.
Other
conceivable 
applications include the potential  detectability of the formation
of
 ``pocket" universes'' in a multiverse.   }  
\end{abstract}

\maketitle
\section{Introduction}
The thought of observing  bursts of \zzs or even  more weakly
interacting \ps from cosmologically early times  raises a number of
interesting issues~\cite{flt}. While  most of the history of the
very early universe is  conventionally described as a succession of
thermal
quasi-equilibrium states, in the more recent universe 
violent phenomena such as neutrino bursts, gamma-ray bursts,
or cosmic ray flares frequently occur. It is perhaps not
excluded
 that   bursts of various kinds   also exist from the earlier
universe.
Particularly if these bursts were to contain weakly
interacting  \psn, the \ps would  be
``messengers''   carrying information from very early epochs. 
Time-of-flight effects, for example, can  provide the
opportunity to determine
the parameters of the  spacetime geometry at the early epoch.

Ignoring for the moment the difficult problem of detectability,
this nevertheless raises the   intriguing question of what is
potentially observable from earlier and earliest epochs, and 
in this note we would like to discuss the in-principle
direct observability of such very early, non-thermal phenomena. 
This is at least 
 of philosophical interest in connection with ideas 
like the ``landscape'' or 
 ``multiverse'' inspired by string/M
theory, or
eternal inflation and quantum gravity implications of
  black hole formation and ``baby universes''~\cite{davies}.  While
these  
events often lead to  new
regions of
 causally disconnected space, it does not seem implausible that
they would leave some trace on our past light cone.
The 
possibility of direct observation of such events 
 --even in principle- moves such ideas out of
  the realm of purely non-physical speculation
into  the world of at least  ``gedanken experiments''.

In general, the more weakly interacting the \pn, the further back
a
burst of some kind can reach us without being lost through
thermalization  by the intervening matter and radiation. While
 the ``look back time'' with photons is  limited  to a few hundred
thousand years after the big bang,  with \zzs we can get to a few
seconds ($\nu_e$) or milliseconds ($\nu_\mu,\nu_\tau$),
 with the yet to be discovered wimps to perhaps 
 picoseconds, and
with gravitons to the epoch of inflation  or to the Planck time.
Because of the general relativistic effects involved with
such early times or high redshifts $z$, a number of interesting
features would be involved with these phenomena, which we would
like
to briefly identify.

\section{Flux dilution, duration, and angular size}\label{flux}
We discuss some features that
might be expected for  bursts emitted at cosmic time $t_{em}$ as
they arrive at
the earth.
We  use the  simplest schematic FRW cosmology; the necessary
quantities are 
briefly reviewed in the appendix.
The
spatial curvature is taken to be zero. 
We concentrate on the case of zero mass or highly
relativistic particles.

\bigskip
{\bf Flux Dilution: } 
The intensity of a burst upon reaching the earth will of course
be greatly reduced by
a geometric flux dilution factor. However, this factor rapidly
approaches a limit at early times.  
In the  flat space result,  that for a  
source  emitting N  particles  the 
number  traversing unit area at a distance $D$ is $N/(4\pi D^2)$,
we
use  
 the $D$ of Eq~\ref{D} (Appendix), so that 
\begin{equation}\label{dil}
{ no.~ crossing~ unit~ area~}=N {1\over 4\pi D^2} \approx N
{1\over 4\pi}({1\over
3t_{now}}
)^2\approx N\cdot 6\times 10^{-59}/{\rm
cm}^2\;~~~~~~~~~~~~~~~~~~~~z>>1.
\end{equation}
The limiting value  for D  is essentially reached by $z\sim
10$ and
thereafter
there is no significant further  reduction as
 z increases.  Thus for these early times, the number of \ps
potentially available for
detection
depends only on N, the number  emitted. We assume isotropic
emission;
Eq~\ref{dil} can be enhanced if there are beaming effects, as with
the jets of galaxies, where the emission  is into less than the
full $4\pi$ of solid angle. 

\bigskip
{\bf Angular size: } $D_H$ is the size of a causally connected
region at $t_{em}$, while the dimension of our backward light cone
is $R(blc)$. Since the burst can at most come from a region of size
$D_H$, the angular dispersion of the burst is limited by 
$D_H/R(blc)$

\begin{equation}\label{disp}
(angular ~dispersion)\approx {(t_{em}/t_{now})^{1/3}\over 1-
(t_{em}/t_{now})^{1/3}}= 1\times 10^{-6}{(t_{em}/s)^{1/3}\over 1-
(1\times 10^{-6})(t_{em}/s)^{1/3}}\;{\rm rad}~~~~~~~~t_{em}>t_{eq}
\end{equation}

\begin{equation}
(angular ~dispersion)\approx{2\over 3 }{(t_{em} t_{rad})^{1/2}\over
t_{now}}=7\times 10^{-9}(t_{em}/s)^{1/2}\;{\rm rad}
~~~~~~~~~~~~~~~~~~~~~t_{em}<t_{eq} 
\end{equation} 

Thus for not too early times, there is a conceivably detectable
angular spread of the burst, if it is emitted from a horizon-sized
region.

{\bf Burst duration :}
The redshift  stretches all time scales. Thus a
\zz burst-producing event at  the \zz decoupling 
time of $\sim 1 s$, if it lasted one ms then, would last a year
upon
reaching us today~\cite{flt}. It would then seem that even
potentially large bursts would be stretched to invisibility.
However, there is a
consideration working the opposite way.  Burst events, representing
a causally connected phenomenon, cannot last longer than $D_H$ or
the age of the universe at the
time they occur, $\sim t_{em}$. This
 argument is similar to the one used in inferring the maximum size
of a
source from
flaring phenomena. Thus
 we would conclude that the maximum duration $\tau_{now}$ of a
burst
upon reaching  us is $\sim t_{em }/a_{em}$ and so for early times
\begin{equation}\label{duration}
\tau_{now}=2t_{em }/a_{em}=2(t_{rad} t_{em})^{1/2}
=9\times 10^{9}\,(t_{em}/s)^{1/2}~{\rm s}~~~~~~~~~~~~~t<t_{eq}
\end{equation}
 
Thus the decreasing time scale wins out over the redshift and the
time scale for burst shortens at early times. At later times, of
course structures involving time scales shorter than the 
Hubble
time can form.
Thus we would expect, as we  go further back, that the bursts first
tend
to lengthen at moderate $z$'s, and then tend to get shorter,
following Eq~\ref{duration}. At the
QCD phase transition, for example, Eq~\ref{duration} gives a burst
duration of  about a year, while at the Planck time
it is
$\sim 10^{-12} s$.

\section{Fluxes} 
A prime question, evidently, for the conceivable observability of
the bursts
is the intensity or rate  to be anticipated.
We proceed on the assumption
that at a given epoch of cosmic time  there is a uniform population
of objects which can emit a certain type of burst with some 
probability during an interval  $d t_{em}$ around  $t_{em}$. Let
this emission probability be described by  $\rho(t_{em})$, a rate
per 
$cm^3 s$ for the burst rate at the epoch $t_{em}$. With $E_{burst}$
the energy of the
burst in the local rest frame at that epoch, bursts in the time
interval
$dt_{em}$ will create a uniform radiation field with  an energy
density

\begin{equation}\label{edens}
d~(energy~ density)_{em}=E_{burst} \rho\, dt_{em} 
\end{equation}
Redshifted to the present with the factor $a_{em}^4$ appropriate to
an energy density of relativistic particles this
becomes

\begin{equation}\label{edensa}
d~(energy~ density)_{now}= a_{em}^4 E_{burst} \rho\, dt_{em} 
\end{equation}
for the present energy density of this radiation.
For relativistic \psn, this uniform energy density
implies an energy flux per $cm^2 s$ at a detector
 \begin{equation}\label{erab}
d~(energy~flux)= {1\over 3}d~(energy~density) = {1\over 3} a_{em}^4
E_{burst} \rho\, dt_{em}.
\end{equation}
 The quantity $\rho$ represents the rate of bursts per unit four-
volume, and it seems most natural to express it in terms  of the
characteristic dimensions at $t_{em}$.
From Eq~\ref{hor}(Appendix), the spatial density  of horizon
sized
volumes  at early times is $({4 \pi\over 3}(2t_{em})^3)^{-1}$. Let
the
probability per
unit time that one such volume produces a burst be proportional to
some
probability  factor $\cal P$ in some time interval. The only
natural
quantity for the time unit is again the Hubble time $1/t_{em}$ so
we write
\begin{equation}\label{prob}
\rho = {1\over{4 \pi\over 3}(2\,t_{em})^4} {\cal P}
\end{equation}
for the density of bursts per unit spacetime volume.
${\cal P}$ is a dimensionless, in general $t_{em}$-dependent
quantity, and presumably small. In  writing this, Eq~\ref{erab}
becomes
\begin{equation}\label{erac}
d~(energy~flux)_{now}=  {1\over 4\pi} a_{em}^4 E_{burst}
{1\over(2\,t_{em})^4} {\cal P} \, dt_{em},
\end{equation}

or, using $a=(t/t_{rad})^{1/2}$ for early times and Eq~\ref{trad}
(Appendix)
\begin{equation}\label{erac1}
d~(energy~flux)_{now}=  {\pi^2\over 90 }  E_{burst}
{T_{now}^4\over (M_{pl}t_{em})^2}  {\cal P} \, dt_{em},
\end{equation}
{\bf Size of bursts:}  The question now arises as to how large
$E_{burst}$ might be. It seems difficult to give a complete answer
to this question without a theory of quantum gravity, particularly
for the gravitational radiation accompanying abrupt changes in
spacetime. However the only obvious quantity and
presumably that which corresponds to the
 maximal  burst is the energy 
within the horizon at $t_{em}$. In such a ``horizon collapse''
  we would  assume that all the energy within the causal horizon is
radiated away, presumably in the form of \ps with  energies higher
than that corresponding to the original thermal plasma. Again, an
estimate of this new energy scale appears to involve quantum
effects, and one simple suggestion would be that it is $M_{pl}.$

 These ``collapses''  might be
possible with
 inhomogeneous universes where there initially were occasional but
large density
variations. In this case the bursts might be the sole evidence of
the ``failed universes''. On the other hand if there was an
inflation and the
universe is highly homogenous, such events would be unlikely.
However they might
 arise in phenomena of the
``baby universe'' or the ``universe in the lab'' type where an
horizon-sized region
undergoes an inflation and some part of the event remains in causal
contact with us. Evidently we must assume that these events are
random and rare in
order to continue to use  simple isotropic-homogeneous
cosmology.

Adopting this energy within the horizon as an estimate, we set
$E_{burst}=E_{horizon}$ in Eq~\ref{erac1}, where from
Eq~\ref{ehoz} we have
$E_{horizon}=M_{pl}^2\, t_{em}$
 and so obtain
the rather simple formula
\begin{equation}\label{erad}
d~(energy~flux)_{now}=  {\pi^2\over 90 }  T_{now}^4 {\cal P}
({dt_{em}\over t_{em}})  \, 
\end{equation}
for the energy flux at the present   from the epoch $t_{em}$.

The resemblance of this result to the intensity of the CMB can be
understood as follows. The normalization in Eq~\ref{prob} to the
horizon volume states that the burst rate is ${\cal
P}dt_{em}/(2t_{em})$ per horizon volume. If we then assume the
energy of the burst is that in a horizon volume or $u_{em}\times
horizon~ volume$, the horizon volume cancels out, leaving the
energy density $u_{em}{\cal P} dt_{em}/(2t_{em})$. The evolution of
the energy density $u_{em}$ to the present gives the present CMB,
and so we obtain
\begin{equation}\label{erd}
d~(energy~flux)_{now}=d~(energy~flux)_{cmb}\,{\cal P}\,
({dt_{em}\over 2\, t_{em}}) \;,
\end{equation}
which is the same as Eq~\ref{erad}.

  For constant $\cal P$ the formula gives an energy flux coming
about equally from all epochs; as we go to
earlier times, the decreasing $E_{horizon}$ is compensated by the
increasing number of emitting regions.
The only dimensional constant explicitly appearing
is $T_{now}$,  the present temperature
of the microwave background, or alternatively  the energy or time
scale  in the radiation dominated epoch via  Eq~\ref{en} or 
Eq~\ref{trad}.  However, various constants will   presumably  be
involved in $\cal P$.

{\bf Energy per burst: }
As for 
 the energy  in an individual, {\it single} burst, Eq~\ref{ehoz}
 (Appendix) red-shifted to the present (with
relativistic
particles) and with the flux dilution factor from Eq \ref{dil}
gives the energy crossing per cm$^2$  as
\begin{equation}\label{eflow}
E_{horizon}^{now}/cm^2=E_{horizon} {a_{em}\over 4\pi D^2}=
{M_{pl}^2
t_{rad}\over 4\pi
(3t_{now})^2}  (t_{em}/t_{rad})^{3/2} =3\times
10^{3}
(t_{em}/s)^{3/2}~{\rm eV/cm^2}\;,
\end{equation}
for the energy crossing unit area  at the detector, from a
single horizon-sized
burst, from $t_{em}<t_{eq}$.

 Thus   an individual, originally
horizon-sized 
burst from around $t_{em}\approx 0.01 s$ would appear in a cm-sized
detector as simply a small-- eV
sized --
but  athermal
signal. ``Athermal'' signifies   that the energy or effective
temperature is larger than that  of the relic distribution of the
particle in question. In addition, the ``burst'' is spread over a
considerable time period, according to Eq~\ref{duration}.

{\bf Burst Rate:} In addition to the energy flux we can also note
the burst
{\it rate} 
corresponding to an assumed $\rho(t_{em})$. The number of events 
from  an interval of cosmic time  $dt_{em}$ and arriving to us
during an interval
$\Delta t$ will be $\rho \, dt_{em}\times (surface~area)\times
\Delta
R $, where  $\Delta R$ is the separation of distance at the epoch
$t_{em}$ leading
to an arrival  time difference $\Delta t$.  The relation is
$\Delta R=a_{em}\Delta t$. Thus since on our backward light cone
$(surface~ area)=4 \pi R^2  $
with $R$  given by Eq~\ref{blca},
 the rate is

\begin{equation}\label{bra}
d(bursts/s)=\rho~ 4\pi R^2 a_{em} dt_{em}, 
\end{equation}

We note that this burst rate, when multiplied by the energy per 
burst, Eq~\ref{eflow}, and using $R=a_{em} D$ yields
Eq~\ref{edensa} for the energy density, checking the consistency of
the formulas.  Inserting Eq~\ref{prob} for $\rho$

\begin{equation}\label{brb}
d(bursts/s)={3\over 16} D^2 {\cal P}\,  ({1\over t _{em}t
_{rad}})^{3/2}\;{dt_{em}\over t _{em} }  =
5\times 10^6 \,{\cal P}\,
{1\over(t_{em}/s)^{3/2}}\;   {dt_{em}\over t _{em} } \, 
{\rm /s}.
\end{equation}
Thus if the intrinsic probability at $t_{em}= 1\, s$ were ${\cal
P}= 10^{-6}$, there would be about one burst per second originating
from this epoch. Originating  from the Planck time there would be
$\sim 10^{65}$ bursts per second with this $\cal P$.
The burst rate is more strongly divergent at small $t_{em}$
than the energy flux since it is not weighted with the 
decreasing energy per burst.

{\bf ``Olber's paradox'':}  We may wonder if  with the very rapidly
increasing
number of independent emitting regions 
 we do not finally produce a kind of  ``Olber's paradox'', where
the amount of energy received  from earliest times diverges. Given
an assumption on the
$t_{em}$ behavior of ${\cal P}(t_{em})$, the total energy received
can be found from Eq~\ref{erad}
\begin{equation}\label{eraa}
(energy~flux)_{now}\propto \int {\cal P}  { dt_{em}\over
t_{em} } \; .
\end{equation}
Thus 
if $\cal P$ is constant  there is a mild Olbers Paradox in the
form of a
logarithmic divergence. With a  cutoff   presumably at the Planck
time  this would give  a further factor $ln (M_{pl} t_{now})\sim
10^2$ in the flux estimate CMB $\times {\cal P}$.
The magnitude and form of ${\cal P}(t_{em})$ is  of course the
great
unknown in our estimates.

{\bf Recent epochs: } The  estimates in this section were all for
$t_{em}$ in the radiation dominated epoch. While horizon-sized
bursts do not seem very likely after $t_{eq}$, where the horizon is
already tens of thousands of years, for completeness we give the
analog
of
Eq~\ref{erad}: 
\begin{equation}\label{erad2}
d~(energy~flux)_{now}= {16\over 81} \rho_{now}\,a_{em}   {\cal P}
\,{dt_{em}\over t_{em}} ~~~~~~~~~~~~~~~~~t_{em} > t_{eq}\, 
\end{equation}
which again follows from  using the Friedman equation, with
$\rho_{now}$  the present matter density. 

\section{``Optical depths''}
 We have, by assumption, taken   the ``medium'' to
be transparent to the
particles of the burst. However as go back to earlier and earlier
times the increasing density of the universe will lead to a greatly
increasing column density and  limit the
``optical depth'' from which a particular particle will be able to
reach us.  To  examine the ``transparency''  we attempt
some rough estimates.

{\bf ``Energy Depth'': } A burst from very early times will travel
through
rapidly varying epochs with a high  but decreasing matter and
radiation density. The effects of this will of course depend very
much on the
nature of the interaction of the burst and this medium, but for
general orientation it is perhaps useful to estimate the ``column
density'' traversed by a burst. Without any specific assumptions as
to the nature of the \ps and their interaction, the only plausible
measure would seem to be the energy/cm$^2$
along the ``line of sight''. Calling this quantity $d$, we
 integrate the energy density along the ``line of sight''.

\begin{equation}\label{dep}
\begin{split}
d&= \int_{t_{em}}^{t_{now}} u dt\approx{\pi^2\over 15}
T_{now}^4(t_{rad})^2\int_{t_{em}}^{\infty} (1/t)^2
dt
=4\times 10^{48} {1\over (t_{em}/s)}~{\rm eV/cm^2}\\
&~~~~~~~~~~~~~~~~~~~~~~~~~~~~~~~~~~~~~~~~~~~~~~~~~~~~~~~=
7\times 10^{15} {1\over (t_{em}/s)}~{\rm gm/cm^2}  
~~~~~~~~~t_{em}<t_{eq}\; .
\end{split}
\end{equation}
At the
Planck time 
 this reaches the rather substantial
value of
$\sim 10^{58}{\rm  gm/cm^2}$.  

{\bf Gravitons: } In terms of specific particles, we can  consider
the
optical depth for  the most  weakly
interacting particle we presume to know about, the graviton. To do
this,
we need  the cross section for the interaction of a graviton with
the
background medium, which just on dimensional
grounds would  be 
\begin{equation} \label{gsig}
\sigma\sim G^2 E\omega\; ,
\end{equation}
with E the  energy of the graviton and
$\omega$ the energy of the background particle.
The mean-free-path for the graviton 
 is then $\lambda= 1/(\rho \sigma)$, with $\rho$ the number
density.
Thus  $\omega \rho$ or the energy density $u$ arises, and
\begin{equation} \label{lambda}
1/\lambda \sim G^2 E u \, .
\end{equation}
To estimate from what ``depth''  a graviton would reach us we
should find that  $t_{em}$ for which 
\begin{equation} \label{depth}
\int^{t_{now}}_{t_{em}}(1/\lambda)dt \approx 1
\end{equation}
If we take a graviton emitted at the Planck time with energy
$M_{pl}$ there is  actually no need
to
explicitly do the integral. It will be seen that (with our usual
neglect of the number of degrees of freedom factor) $u$ and all 
other factors involve only $M_{pl}$. Since there is no other
constant involved for the dimensionless integral Eq~\ref{depth},
we will obtain
$\int_{t_{em}}^{t_{now}}(1/\lambda)dt\approx
\int_{t_{pl}}^{\infty}(1/\lambda)dt\sim 1$. Hence we conclude
that an $E\sim M_{pl}$ graviton emitted at the Planck time can just
about
reach us. Lower energy or more recently emitted particles will
naturally find it easier to reach us. On the other hand more
strongly interacting
or perhaps higher energy transplanckian single particles will be
absorbed,
if their interaction is as in Eq~\ref{gsig}.

In general, since the burst particles are presumed to be emitted
with
energies at or above the \t of the ambient medium, they should
travel freely at times around  or somewhat later than that for the
decoupling of that species. Thus taking this time,
 for neutrinos we anticipate
$t_{em}>10^{-3} s$ and  for a wimp $\chi$,
$t_{em}>10^{-11}(m_{\chi}/100
GeV)^{-3/2} s$, while with
gravitons we can go back to $t_{pl}=0.6\times 10^{-43}s$.

{ \bf Recent times:}  It is interesting to note that the $d$
of Eq~\ref{dep}
reaches about a thousand grams, the depth of the earth's
atmosphere,
around
$t_{em}\approx t_{eq}=2\times 10^{12}s$. This suggest that also the
usual components of cosmic rays, protons and photons, can
 reach us from  epochs before the present one. More specifically,
now working with the baryon component and
taking the 
baryon
density  at 0.04 of the critical
density, we have at present an average baryon density of
$\rho_{now}=4\times 10^{-31} gm/cm^3$. Scaling 
with $a$, Eq~\ref{teq} (Appendix), we have a column density of
baryons for a \p
emitted at $t_{em}>t_{eq}$
\begin{equation}\label{byn}
d_{baryon}\approx 4\times 10^{-31} {\rm gm/cm^3}
\int_{t_{now}}^{t_{em}}
(1/a^3) dt= 2\times
10^{15}({1\over (t_{em}/s)}-{1\over (t_{now}/s)})~{\rm gm/cm^2}\;,
\end{equation}
which yields a thousand  grams at $t_{em}\approx t_{eq}$.
Thus  matter-radiation
equality is around the  ``horizon''  for
ordinary cosmic rays and  sets    the limit where
 weakly interacting \ps become necessary for a burst to reach
us~\cite{topdown}.

{\bf  Transplanckian epoch:}
At the opposite end of the time scale, we may speculate concerning
the existence of direct signals  from the quantum gravity or
transplanckian epoch. This leads us to a curious question: is it
possible to have an arbitrarily weakly interacting \pn?

The question arises since according to the arguments around
Eq~\ref{depth} 
 it might appear that we need only postulate the existence of a \p
whose interactions are weaker than that of the graviton, less than
Eq~\ref{gsig}, in order for the \p to reach us from ``before'' the
Planck time. On  second thoughts, however, this is not possible
within the framework of conventional \p physics. Gravity couples 
universally to all forms of energy, and so any particle  with mass
or energy  must inevitably have at least an  interaction on the
order of $\sigma\sim G^2E^2_{cm}$. Thus it would appear we are
forced to conclude that transplanckian signals cannot reach us. 

 A way around this seemingly unavoidable conclusion, suggested
to us by V. I. Zakharov, might be offered  by string theory or
other models where the graviton is imbedded in a complex 
representing many \psn,  like the ``string''. The whole string or
complex could  be more weakly interacting than the graviton itself.
In this way the object might  be able to ``get out'', even though
viewed as individual
components this is not possible.   This is an interesting  line of
thought, but goes  beyond the straightforward \p- based
ideas we  use here. 

\section{characteristics of the signal}
 We discuss some  aspects of the signal following from the above
considerations. In
doing so we are not, unfortunately, able to deal with the most
important point, the
\ddn method. This promises to be very difficult since by assumption
we deal
with very weakly interacting particles. Nevertheless some aspects
of the
presumed signal are novel and interesting.

For not too early times, the burst will qualitatively be like those
familiar
from, say, supernova \zz bursts or gamma-ray bursts.
As we go to earlier times, and if $\cal P$ is not extremely small,
we  
 likely have the situation of  a high rate of low energy bursts--a
steady ``rain''
of weak signals. Since the particles are by assumption  originally
more
energetic than that of the ambient  plasma at the time of emission,
the
individual quanta should be more energetic than those of the
present CMB and
so not of extremely low energy. Thus, unless the \dd is of very
large area
or volume, the burst appears essentially as isolated single quanta.
On the other hand, the fact that  we  indeed deal with a burst
 and not a random signal offers 
 the possibility of
coincidences between separated \ddsn , should a sufficiently
efficient
\ddn method exist.

 There are thus two qualitatively different situations according to
whether the burst rate or equivalently the time between bursts is
large or small compared to the length of the bursts. If the bursts
arrive singly there is the possibility in principle of extracting 
some kind of coincidence or coherence information, perhaps from
multiple \dds. On the other hand if many bursts  overlap in time
this is not possible; unless, that is, the detectors have good
angular resolution and can identify individual bursts by 
direction.

Finally, we note that a most interesting question concerning the
signals
would be  directional isotropy --- or the lack of it. The
directional isotropy is one of the most striking feature of the
CMB, and it would
be of great interest to know if this is still the case for 
radiation from earlier epochs. 

\section{Degrees of freedom, inflation, dark energy} We have made
a number of simplifying assumptions, including considering only one
radiation field ($g(T)=2$)~\cite{kt}, and the neglect of inflation
and dark energy, if they exist.
Concerning dark energy or a late-time expansion, its effects are 
anticipated to only become
relevant in recent epochs where, as for  the observations of   type
1a
supernovae, they are small and should not have significant effects
on the above considerations. 
Concerning $g(T)$, the entropy conservation
condition for the interacting plasma $(Ta)^3=constant$ or  $T\sim
1/a$ becomes $T\sim 1/(g^{1/3}a)$\cite{kta}, so that instead of
$u\sim 1/a^4$ for the energy
density
 one has $u\sim 1/(g^{1/3} a)^4$.  If one
makes an adiabatic assumption so that  $g$ may be treated as a
constant
in the various equations,  $g$  appears in the form of small
fractional
powers and  our general estimates will not be qualitatively
altered. Interestingly the effects cancel in  Eq~\ref{erad}, which 
remains unchanged.

If we adopt the inflationary scenario, the previous considerations 
can still be used, with the proviso  that they be used
after the end of inflation.
Any bursts emitted before inflation will of course be inflated away, 
shielding the quantum gravity epoch from view. The associated
smoothing of
the energy density, leaving only the $\sim 10^{-5}$
fluctuations
later seen at recombination, would tend to eliminate potentially
more violent fluctuations. It thus  appears that the absence of
very early bursts is a prediction of inflation. On the other hand
the possibility of inflationary phenomena like ``baby universes''
might still leave burst-like traces in our history. Although a more
detailed study of such occurrences would be interesting,
qualitatively the scale for such phenomena would appear  to be
again
the horizon, as in our discussion.

\section{summary}
 We have suggested that, in principle, it would be possible to
detect
bursts  of weakly interacting
particles 
such as \zzs or
even more weakly interacting particles such as wimps and gravitons
from the very early universe. This
would
offer  a much
deeper  ``look
back time'' to earlier epochs  than is possible with photons. We
have
considered some of the
issues related to  the existence of such bursts  and their        
   phenomenology.  The
existence of such
bursts would make the
observation of  phenomena associated with very early times in
cosmology,   or perhaps even the transplanckian epoch, at least
conceptually possible. For the latter it seems necessary to posit
the existence of
complexes more weakly interacting than the graviton. Possible
applications include the conceivable detectability of the formation
of
 ``pocket" universes'' in a multiverse as well as similar
phenomena.
In this application  we  should  stress that we do not wish to
 suggest in any way
 that there can be contact
or information transferred  from ``other universes''. Rather the
bursts we
discuss would be edge effects or transition  phenomena taking place
during the
short periods of formation of the new region of spacetime. 
\vskip2.5cm
\centerline{\bf APPENDIX:~~Description of the early universe }
\vskip0.2cm

We briefly review the basic quantities used in our simple
cosmological model (units
$\hbar, c, k_{Boltzmann}
=1$.):

{\bf Scale Factor:} We use the scale factor
$a(t)$, which at present is $a_{now}=1$ . For times after matter-
radiation  equality  at $t_{eq}$, one has $a\sim t^{2/3}$ 
and before $t_{eq}$, in the radiation dominated epoch, $a\sim
t^{1/2}$
\begin{equation}\label{teq}
a=(t/t_{now})^{2/3}~~~~~~~t>t_{eq};~~~~~~~~~~~~~~~~~~~~~~~~~~~~   
a=(t/t_{eq})^{1/2}~
a_{eq}=(t/t_{rad})^{1/2}~~~~~~~~t<t_{eq};       
\end{equation}
where  we introduce the quantity 
$t_{rad}=t_{eq}/a_{eq}^2$. 
We thus  have the  parameters
\begin{equation} \label{param}
t_{eq}=2\times 10^{12}s~~~~~~~~~~~ t_{now}=4\times 10^{17}s~
~~~~~~~~~~~t_{rad}=2\times10^{19} s ~~~~~~~~~~a_{eq}=3\times
10^{-4}
 \; .
\end{equation}

 We  use $t_{em}$ to refer to the time of emission of the
burst. The redshift of an energy or frequency from $t_{em}$ to the
present  is $E_{now}/E_{em}=a_{em}=1/(1+z)$ or  $\approx 1/z$ for
large $z$.

\bigskip

{\bf Energetics:} Also,  we need the energetic quantities   energy
density,  \t and
the energy in a horizon volume at early times. We
 work with a simplified
universe with only one radiation field, so that also for the \t
\begin{equation} \label{at}
T_{now}/T(t)=a(t)\; ,
\end{equation}
where $T$ is the \t of the radiation field and in particular 
$T_{now}$ is the present \t of the CMB.

For the radiation-dominated era we thus have for the energy density
\begin{equation}\label{en}
u= {\pi^2\over 15} (T_{now} /a) ^4 ~~~~~~~~~~~~~~~~~~t<t_{eq}\;.
\end{equation}
With more fields present at high \tn, the effective $T_{now}$ in
Eq~\ref{en} will
change somewhat, however for our present
semi-quantitative
purposes we shall take it to be the constant 
$T_{now}\approx 3^0K\approx 2.5\times 10^{-4} eV$.

We note that the Friedman equation in the radiation dominated era 
connects $T_{now}$ from Eq~\ref{en} and $t_{rad}$ in Eq~\ref{teq}:
\begin{equation}\label{trad}
T^2_{now} t_{rad}=\sqrt{45\over 32 \pi^3}  M_{pl}\;. 
\end{equation}

{\bf Distances: }To find the energy within the horizon at some
early time $t_{em}$, we first  need the size of the horizon, which
is given by  $D_H=a_{em}~ \int^{t_{em}}_0(1/a)dt$, leading to
\cite{kt}
\begin{equation}\label{hor}
D_H=2
t_{em}~~~~~~~~t_{em}<t_{eq};~~~~~~~~~~~~~~~~~~~~~~~~~~~~~~~~~~~~~
~D_H= 3 t_{em}~~~~~~~~~t_{em}>t_{eq}\;.
\end{equation}

The energy within the horizon at early times now follows 
by   multiplying  $u$ in the Friedman equation by the volume
factor $\sim D_H^3$ to obtain
\begin{equation} \label{ehoz}
E_{horizon}=u{4\pi\over
3}(2\,t_{em})^3=M_{pl}^2\,t_{em}=M_{pl}\,(t_{em}/t_{pl})~~~~~~~~~
~~~~~~~~~~t_{em}<t_{eq}
\end{equation}
using $a\sim t^{1/2}$ and the Planck time
$t_{pl}=1/M_{pl}=0.55\times 10^{-43}s$. For later times the same
argument leads to essentially the same result
\begin{equation} \label{ehoza}
E_{horizon}=6M_{pl}^2\,t_{em}~~~~~~~~~~~~~~~~~~~~~~~~~~~~~~~~~~
~~~~~¬~~~~~t_{em}>t_{eq}\;.
\end{equation}
We  also need the distance 
$D=\int_{t_{em}}^{t_{now}}(1/a)dt$, which is given
essentially by the
matter dominated epoch:
\begin{equation}\label{D} 
D\approx\int_{t_{em}}^{t_{now}}(1/a)dt=3t_{now}[1-(t_{em}/t_{now}
)^{1/3
}]\approx 3
t_{now}\;.
\end{equation}

 $D$ enters into the radius of our backward light cone,
$R(blc)$ at $t_{em}$ via  $R(blc)=a(t_{em})
\int_{t_{em}}^{t_{now}}(1/a)dt$. 
so that
\begin{equation}\label{blc}
R(blc)= a_{em}D=3t_{now}[(t_{em}/t_{now})^{2/3}-(t_{em}/t_{now})]=
3t_{em}[(t_{now}/t_{em})^{1/3}
-1]~~~~¬~~~~~~t_{em}>t_{eq}
\end{equation}
and at early times 

\begin{equation}\label{blca}
R(blc)=a_{em}D=3t_{now}(t_{em}/t_{rad})^{1/2}
~~~~~~~~~~~~~~~~~~~~~~~~~~~t_{em}<t_{eq}\;.
\end{equation}


\begin{thebibliography}{00}

\bibitem{flt} L. Stodolsky, Physics Letters {\bf B 473},
{61}{2000}, and Looking Back with Neutrinos,
 {\it Proceedings, Carolina Symposium on Neutrino Physics}, 
Columbia, March 2000; Eds. J. Bahcall, W. Haxton, K. Kubodera, and
C. Poole; World Scientific, Singapore. 
 astro-ph/0006384. 


\bibitem{davies}Testability issues connected with such ideas
 have recently been reviewed by P. Davies, astro-ph/0602420, G.
Ellis
astro-ph/0602280, or W. Stoeger astro-ph/0602356.    

\bibitem{kt} E. Kolb and M. Turner, {\it The Early Universe},
 Addison-Wesley, (1990) pg 82.
\bibitem{kta} Kolb and Turner, {\it ibid} Eq~3.77.

\bibitem{comp} We note this does not depend on assuming a
thermodynamic character for the energy density in question. If the
\ps in question are massive the shift to the present time is more
complicated.


\bibitem{topdown} Cosmic rays originating before recombination are
a 
possibility in  ``top-down'' models, see
 P. Bhattacharjee and G. Sigl, Physics Reports  {\bf 327}, 109,
(2000).
We do not consider very high energy cosmic rays where effects of
scattering
on the CMB or infrared background become important.



\end{thebibliography}
\end{document}